\documentclass[twocolumn]{aastex6}
\usepackage{graphicx}
\usepackage{epstopdf}
\usepackage{natbib}

\begin{document}

\title{Plasma heating in solar microflares: statistics and analysis}
\author{Kirichenko A.S. and Bogachev S.A.}
\affil{P.N. Lebedev Physical Institute of the Russian Academy of Sciences \\
Moscow, 119991, Russia}

\begin{abstract}

In this paper, we present the results of an analysis of 481 weak solar flares, from A0.01 to the B \textit{GOES} class, that were observed during the period of extremely low solar activity from 2009 April to July. For all flares we measured the temperature of the plasma in the isothermal and two-temperature approximations and tried to fit its relationship with the X-ray class using exponential and power-law functions. We found that the whole temperature distribution in the range from A0.01 to X-class cannot be fit by one exponential function. The fitting for weak flares below A1.0 is significantly steeper than that for medium and large flares. The power-law approximation seems to be more reliable: the corresponding functions were found to be in good agreement with experimental data both for microflares and for normal flares. Our study predicts that the evidence of plasma heating can be found in flares starting from the A0.0002 X-ray class. Weaker events presumably cannot heat the surrounding plasma. We also estimated emission measures for all flares studied and the thermal energy for 113 events.

\end{abstract}


\section{Introduction} \label{sec:intro}

Microflares are flaring events in which the total energy lies in the range of $10^{27}-10^{30}$ erg. In general, in microflares we can observe the same active phenomena as in ordinary flares but on smaller scales: plasma heating, acceleration of charged particles, and even ejections of coronal masses \citep{Kirichenko2013}. The process of plasma heating during solar flares of different classes was previously studied in several works. One of the earliest studies was performed by \citet{Garcia1992}, who calculated the emission measure of 710 solar flares of M- and X-class based on GOES data, and found a correlation between the X-ray class of a flare and its emission measure.\

\citet{Feldman1996} analyzed the large sample of flares comprised of data from \textit{GOES} and from the Bragg Crystal Spectrometer (BCS) on board \textit{Yohkoh}. They selected 868 flares from the A2 to X2 \textit{GOES} classes and found their temperatures using BCS data, and calculated their emission measures by convolution with the corresponding \textit{GOES} data. The study revealed a logarithmic relationship between \textit{GOES} class of a flare and the temperature of heated plasma, and a power law relationship between the GOES class and emission measure. Both of these parameters grow with increasing \textit{GOES} class.\

\citet{Battaglia2005} performed a similar analysis for a sample of 85 flares with X-ray classes from B1 to M6, and also obtained the correlation between the X-ray class of a flare and its plasma temperature (at the time of the maximum of hard X-ray emission, measured using RHESSI data). Due to their relationship, the plasma temperature in small solar flares was found to be higher compared to the results of \citet{Feldman1996}. This can be explained by the fact that \citet{Battaglia2005} and \citet{Feldman1996} measured the temperature at different moments of time: during the maximum of hard X-ray (HXR) flux in the first paper, and during the soft X-ray (SXR) maximum in the second. The difference may be explained by the fact that the temperature at the maximum of HXR flux is higher than that at the maximum of the SXR flux. Moreover, the temperatures measured from \textit{RHESSI} are always significantly higher than the temperatures obtained from \textit{GOES} .\

\citet{Hannah2008} analyzed 25,705 flares from classes A to C based on \textit{RHESSI} data, and defined a median temperature and emission measure for these events as $~13$ MK and $3\times10^{46}$cm$^{-3}$ respectively. They also determined the temperature and emission measure as a function of the \textit{GOES} class -- the relations are logarithmic and power law, respectively. The results were close to \citet{Feldman1996} and \citet{Battaglia2005}.\

\citet{Li2012} considered 1843 flares using \textit{RHESSI} and \textit{GOES} data and determined their temperatures at the moment of maximum SXR flux. They also studied the electron temperature and concentration as a function of \textit{GOES} class. Both have power-law relationships.\

The most extensive sample of flares (consisting of more than 50,000 events from B- to X-class) was studied by \citet{Ryan2012} using the data of \textit{GOES}. They developed the TEBBS method, which allowed them to improve the background subtraction and calculation of the temperature and emission measure of flaring plasma. In contrast to other works, they derived the maximum temperature, maximum emission measure, and maximum flux for each flare. They found that the emission measure depends on the \textit{GOES} X-ray flux as a power law, while the temperature grows as a logarithm. The results were principally comparable to previous studies but flares of a given \textit{GOES} class in \citet{Ryan2012} had lower peak temperatures and higher peak emission measures than previously reported.\

\citet{Caspi2014} analyzed 37 strong flares of M- and X-class with \textit{RHESSI} data and among other results determined their temperature and maximum thermal energy as a function of \textit{GOES} X-ray class. They obtained logarithmic and power-law relations for temperature and emission measure as a function of X-ray flux.\

In general, all the previously reported results demonstrate an obvious correlation between the X-ray class of a flare and its temperature and emission measure. However, most of the results were obtained for ordinary flares (classes C--X) and for a small amount of flares of A- and B-class. There is still no information on whether the relationships revealed for A- to X-flares continue to be valid for more faint events in the range of A0.01 -- A1.0. Simple extrapolation of earlier results to the region of microflares predicts that flares of A0.01--A1.0 classes cannot heat surrounding plasma to detectable temperatures. This is in obvious contradiction with the observations, which leads to the idea that plasma temperature and emission measure in solar microflares depend on the  X-ray class of a flare by some other way than what were found for medium and strong events.\

This work presents a statistical analysis of the thermal properties for solar microflares, including extremely weak events up to the A0.01 \textit{GOES} level. For all these events, we calculated the temperature, emission measure, thermal energy, and electron concentration of plasma at the SXR flux maximum in one-temperature and two-temperature approximations. We compare our data with previous results obtained in the A--X range.\

\section{Data and processing} \label{sec:data}
\subsection{Data}
In this paper, we used data from two X-ray instruments operated in 2009 (during the minimum of solar activity) on board the \textit{CORONAS-Photon} spacecraft: SphinX \citep{Sylwester2008}, and MISH \citep{Kuzin2009}. SphinX is an X-ray spectrophotometer, which provided high-precision spectra of solar flares in the energy range of 0.5--15 keV. The significant advantage of the instrument was its high sensitivity: SphinX was able to register X-ray emission for flares of class A0.1 and lower. The second instrument, MISH, is an imaging Bragg telescope, which provided monochromatic images of the Sun in the resonance doublet Mg XII 8.419 and 8.425~\AA\ with a spatial resolution of about 4 arcsec. In order to produce an appreciable emission of these lines, the coronal plasma should have a temperature of about 4 MK or higher. Therefore, the images from MISH reliably indicate the locations of high-temperature plasma in the corona.\

The \textit{CORONAS-Photon} spacecraft operated in orbit from February to November of 2009 during a period of extremely low solar activity, which was favorable for the observation of weak flare events. For this study, we selected only the data from 2009 April to July due to the especially low level of solar activity and good cadence of MISH data during this time. Using the SphinX events catalog we selected 601 weak flares. After preliminary examination their number decreased to 481 due to instrumental and other problems with the rest of the events. These 481 flares were then studied in detail.\

\subsection{Data processing}
\begin{figure}
\centering
  \includegraphics[]{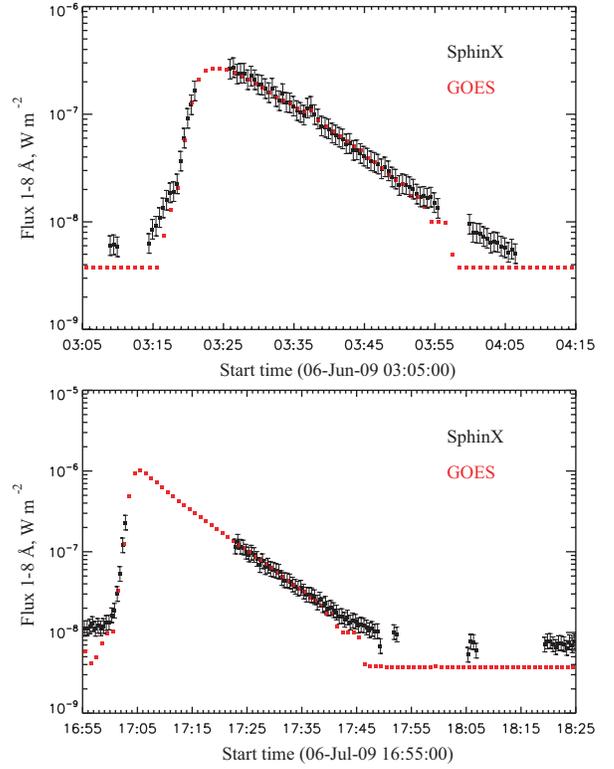}
%    \resizebox{\hsize}{!}{\includegraphics{figs/feld_t_2_hot_good_points_log.eps}}
     \caption{Comparison of \textit{GOES} and calibrated SphinX data.}
     \label{goes_sphinx}
\end{figure}

\begin{figure}
\centering
  \includegraphics[]{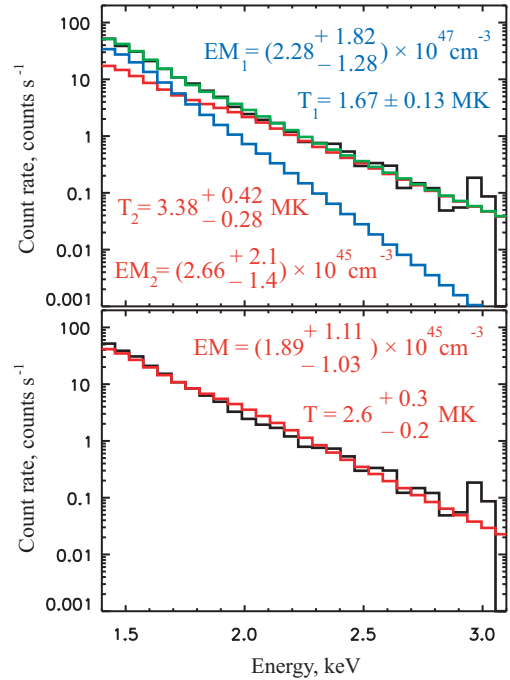}
%    \resizebox{\hsize}{!}{\includegraphics{figs/feld_t_2_hot_good_points_log.eps}}
     \caption{Example of the microflare spectrum fitted by two thermal components (upper panel) and one thermal component (bottom panel).}
     \label{spectra_2t}
\end{figure}

For each selected flare we calculated the electron temperature of plasma \textit{T} and its emission measure EM. For 113 events, for which we had a MISH image within 1 minute from the SXR maximum, we also calculated volume \textit{V}, electron concentration $n_{e}$, and the thermal energy $E_{\rm th}$.The electron temperature and emission measure were obtained from the SphinX spectra by fitting them with one (isothermal model) and two (2T model) thermal components (examples of 1T and 2T fitting of spectrum are shown in the Figure~\ref{spectra_2t}). The SXR flux was calculated on the electron temperature and emission measure derived from spectra fits. Actually, the flux on SphinX data is several times higher than that on the data of other instruments such as, for example, \textit{GOES} or \textit{RHESSI} \citep{Mrozek2012}. \cite{Mrozek2012} found that this difference is due to the shift of emission measure since the temperature values are close. In our work, we performed the data calibration between \textit{GOES} and SphinX to take into account this issue. We used the 1 minute \textit{GOES} X-ray flux in the range 1--8~\AA\ and the corresponding spectra of SphinX for the period from April to August of 2009. The X-ray flux on SphinX data was calculated on the temperature and emission measure derived from SphinX spectra in the isothermal approximation.  The correlation between SphinX and \textit{GOES} X-ray flux in the range 1--8~\AA\ is shown in the Figure~\ref{calibration}. The full set of data contains 7680 points. The relation is linear in a double logarithmic scale for the \textit{GOES} flux above about the A2.0 level, but under this value the \textit{GOES} flux drops dramatically. We believe this is due to the border effects that arise near the \textit{GOES} registration sensitivity threshold. Thus, we considered only points with a \textit{GOES} flux above the A2.0 level (4368 points totally) to obtain the relation between \textit{GOES} ($G$) and SphinX ($S$) flux: 
\begin{equation}
lg G = (1.11\pm0.55)+(1.24\pm0.08)lg S.
\label{calibration}
\end{equation}
Examples of applying this relation to the SphinX data for two flares are in Figure~\ref{goes_sphinx}. The red color squares are \textit{GOES} data and the black squares are SphinX flux. At the start and at the end of both events, the consequences of decreasing the \textit{GOES} efficiency near its sensitivity threshold are clear. We used the relation between SphinX flux before and after correction to remove the overestimation of the emission measure derived from the spectra. All the fluxes and emission measures we use in our work are after correction.\
\begin{figure}
\centering
  \includegraphics[]{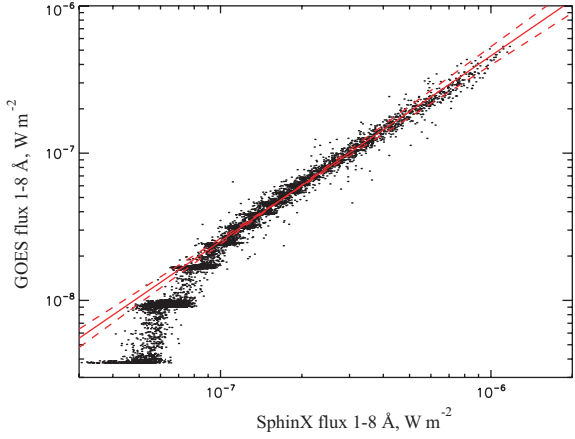}
%    \resizebox{\hsize}{!}{\includegraphics{figs/feld_t_2_hot_good_points_log.eps}}
     \caption{Correlation between \textit{GOES} and SphinX SXR flux in the range 1--8\AA.}
     \label{calibration}
\end{figure}

\begin{figure*}
  \resizebox{\hsize}{!}{\includegraphics{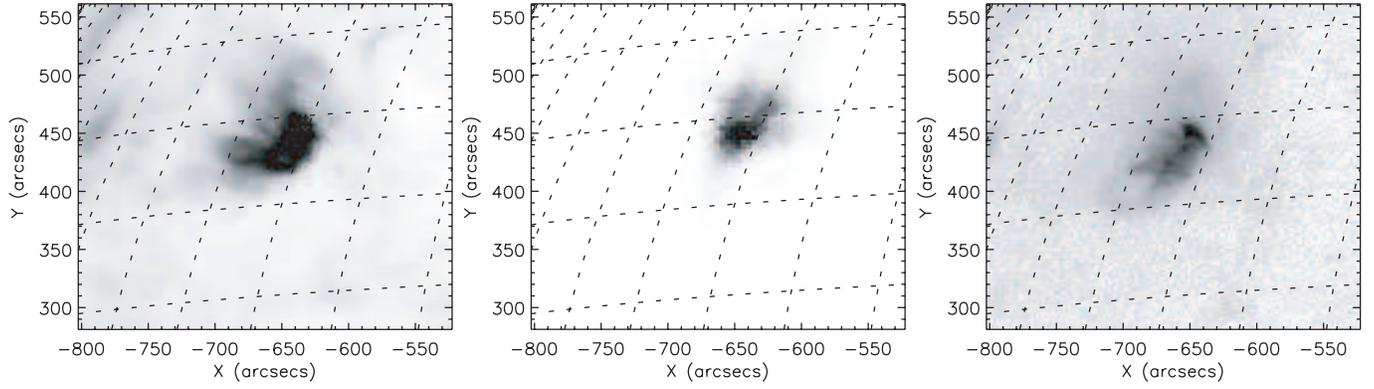}}
  \caption{Images of flaring region in the data of three telescopes: left -- EUV telescope FET/TESIS at 132\AA; middle -- MISH/TESIS; right -- XRT/\textit{Hinode}.
  }
  \label{source}
\end{figure*}

The volume of the emission source was estimated with MISH images using the following procedure: we calculated the signal distribution in the part of the image that contains the X-ray source and determined the background level as the most frequent mean of the distribution. All data that exceeded this level by more than $3\sigma$ were considered useful signals. An example of a flaring region image is in the Figure~\ref{source}. To show the difference between the size of the source in several spectral regions we demonstrate not only MISH data (the middle panel in the figure) but also the data of the EUV telescope FET/TESIS at 132\AA\ and XRT/\textit{Hinode} with Al poly filter. Since MISH took images in two lines, significant parts of the sources are extended along the dispersion direction. To avoid the overestimation of the volume we derived the size of the source in the orthogonal plane to the dispersion direction. We assumed sources to be spheres and calculated their volumes as $V=(4/3)\pi(R)^{3}$. The electron concentration was estimated as $n_{e}=\sqrt{EM/V}$ and the thermal energy was estimated as $E_{th}=3n_{e}k_{B}TV$ where $k_{\rm B}$ is the Boltzman constant.\
%As one can see, hot emission source on MISH image is 'U'-shape due to the aberration, therefore we considered only bright core of the sources.
\section{Results} \label{sec:res}
\subsection{Temperature and emission measure}
\begin{figure*}
\newpage
\centering
   \includegraphics[width=17cm]{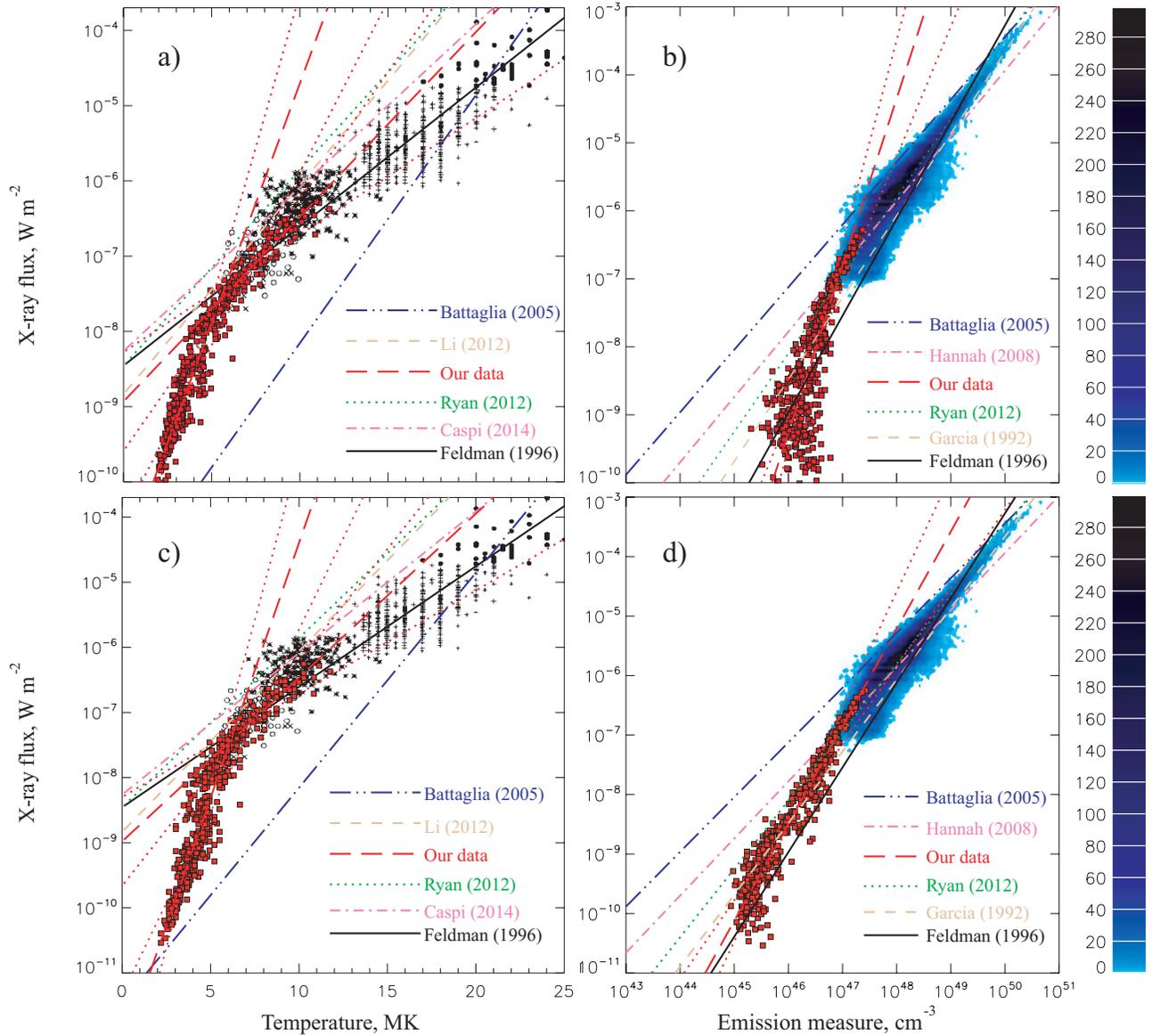}
     \caption{X-ray flux in the range 1--8~\AA\ as a function of plasma temperature and emission measure. The upper panels correspond to the isothermal model and the bottom panels panels correspond to the 2T model. The red coloring marks our data: red squares are the experimental points and the red dashed line is a fit onto them. The black characters are the data of \citet{Feldman1996}. The blue two-dimensional histograms are the data of \citet{Ryan2012}.}
     \label{all}
\end{figure*}
The temperature and emission measure at the moment of the flare's maximum are shown in Figure~\ref{all} as a function of peak flare flux in the range 1--8~\AA. Both \textit{T} and EM were calculated by two approximations: 1T and 2T. Panels a and b correspond to the isothermal model, and panels c and d correspond to the 2T model. Our data are plotted with red squares. In the same figure, we have plotted previously obtained data for ordinary flares from: \citet{Garcia1992}, \citet{Feldman1996}, \citet{Battaglia2005}, \citet{Hannah2008}, \citet{Ryan2012}, \citet{Li2012}, and \citet{Caspi2014}. The full list is displayed in the legend of the plot.\
Normally, in the works mentioned above the authors revealed a linear approximation between the plasma temperature \textit{T} and the logarithm of X-ray flux:
\begin{equation}
log_{10}~PFF=a+b~T.
\label{exponen}
\end{equation}
If we add the data for flares of the A0.01--A1.0 classes to the plot, the relation between X-ray flux and plasma temperature becomes more complicated. The whole data set can be roughly divided into 2 parts with different inclinations -- below and above the A1.0 level. The left part of the curve, which is related to microflares, becomes steeper than the right part, which is related to normal flares. Every part can be approximated by the function (\ref{exponen}), but with its own coefficients a and b as listed in the Table~\ref{table1} (panel I) for 1T and 2T approximations. To estimate the accuracy of the approximations we used the linear (LCC) and Kendall ($\tau$) correlation coefficients. The minimum values of the LCC and $\tau$ for the $log_{10}~PFF(T)$ relation were found to be 0.79 and 0.63, respectively, for the 1T and 2T models, which implies a statistically significant correlation between fits and experimental data.\
\begin{figure*}
\centering
  \includegraphics[]{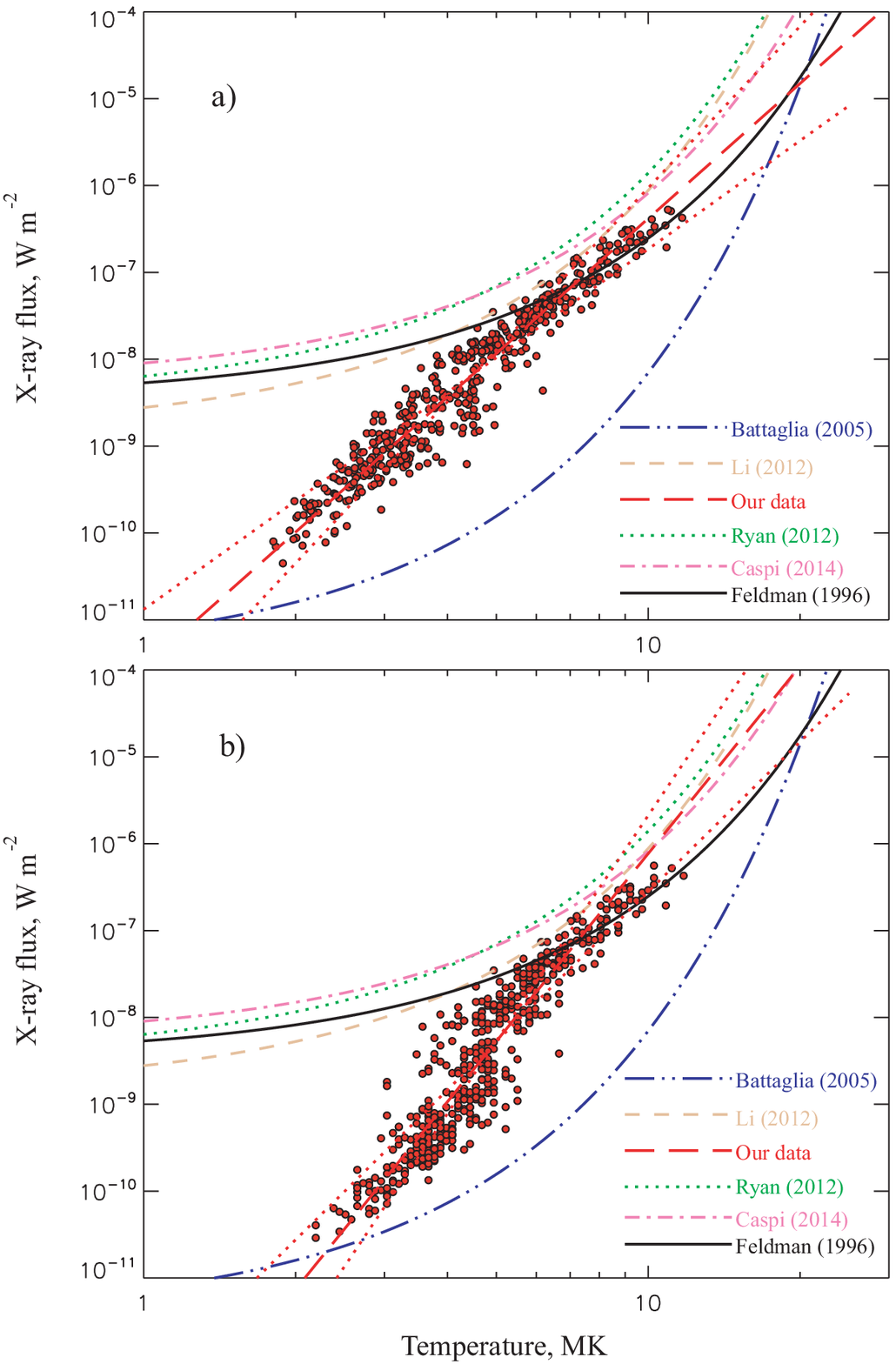}
     \caption{X-ray flux of solar flares as a function of temperature in logarithmic scale. The top panel corresponds to the isothermal model and the bottom corresponds to the 2T model. The red circles and the red dashed line are the experimental points and the fit onto our data. The remaining lines correspond to the fits from other works.}
     \label{log}
\end{figure*}
The relation between X-ray flux and emission measure for the isothermal and 2T models may be fitted by one power-law function without any breaks. The coefficients a and b for these approximations are presented in panel II of the Table~\ref{table1}. The correlation coefficients for both models are high enough to consider the correlations between experimental data and their fits to be statistically significant. For the flares of the A1.0 class and higher, both models (1T and 2T) give almost identical results. The difference between them can be found only for the events below the A1.0 class; the emission measure for weak events in the isothermal approximation is higher than that in the 2T model.\

To fit the dependence PFF$(T)$ with one joint function, we approximated this relation using the next law:
\begin{equation}
log_{10}~PFF=a+b~log_{10}~T.
\label{power_law}
\end{equation}
The results of approximation for the isothermal and 2T models are shown in Figure~\ref{log}. The LCC and $\tau$ are 0.96 and 0.84 for isothermal model approximation, and 0.9 and 0.8 for the 2T model (Panel III in the Table~\ref{table1}). Figure~\ref{log} demonstrates that power-law function (\ref{power_law}) fits the relation PFF$(T)$ well in a wide range of flare classes from A0.01 to B, and higher. For events for the A- and B-classes, the function (\ref{power_law}) is very close to the relationship obtained by \citet{Feldman1996}. However, for microflares below the A-class the plasma temperature is significantly higher than what would be expected from the majority of previous works, except for \citet{Battaglia2005}.\

\begin{figure*}
\centering
   \includegraphics{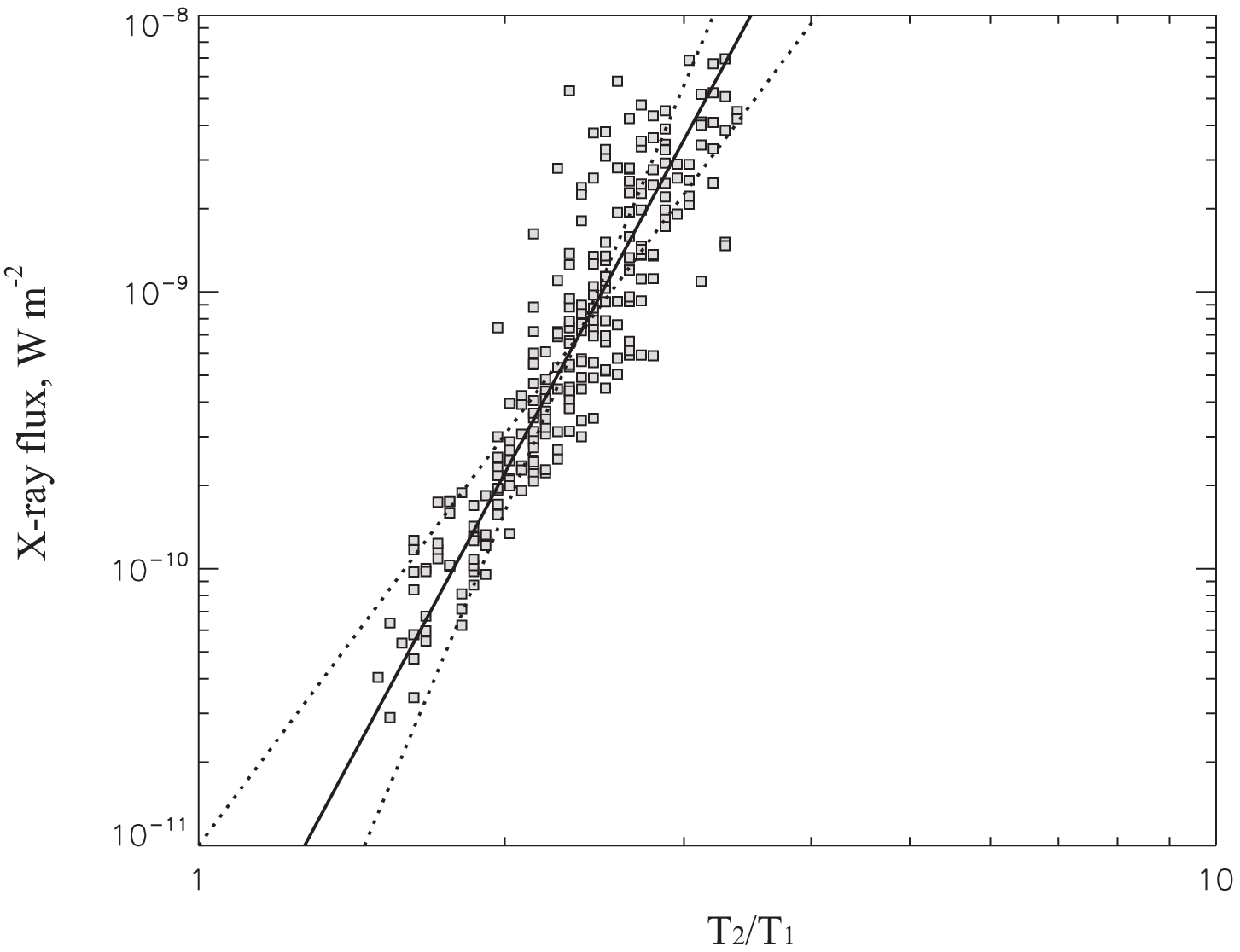}
     \caption{X-ray flux of microflares as a function of the ratio of flaring and surrounding plasma temperatures on a logarithmic scale.}
     \label{t2/t1}
\end{figure*}
The important question is: what is the minimal X-ray class of a flare that is able to heat the plasma to a temperature higher than the surrounding plasma? To answer this question, we calculated temperatures \textit{T}$_{1}$ (cold component) and \textit{T}$_{2}$ (hot component) in the 2T approximation for every flare in our data set. We identified \textit{T}$_{1}$ with the temperature of the surrounding plasma and \textit{T}$_{2}$ with the temperature of the plasma heated during the flare. If the ratio \textit{T}$_{2}$/\textit{T}$_{1}$ decreases to a value of 1, it means that the flare does not heat the plasma (or we cannot detect such heating due to the low level of emission measure). In Figure~\ref{t2/t1} we demonstrate the relationship \textit{T}$_{2}$/\textit{T}$_{1}$ as a function of X-ray flux (the corresponding fitting results are shown in Panel IV of Table~\ref{table1}). We used the data only for microflares of the A-class and lower, because stronger events are better described by the 1T model and comparing the components in the 2T model is incorrect for them. The X-ray flux corresponding to the point \textit{T}$_{2}$=\textit{T}$_{1}$  is $10^{-11.71\pm0.77}=1.9\times10^{-12}$~Wm$^{-2}$ (~10$^{-4}$ of the flare level A1.0). The corresponding temperature of plasma in the power-law approximation (\ref{power_law}) is $T_{2}=1.66\pm0.34$ MK.\

\begin{figure*}
  \resizebox{\hsize}{!}{\includegraphics{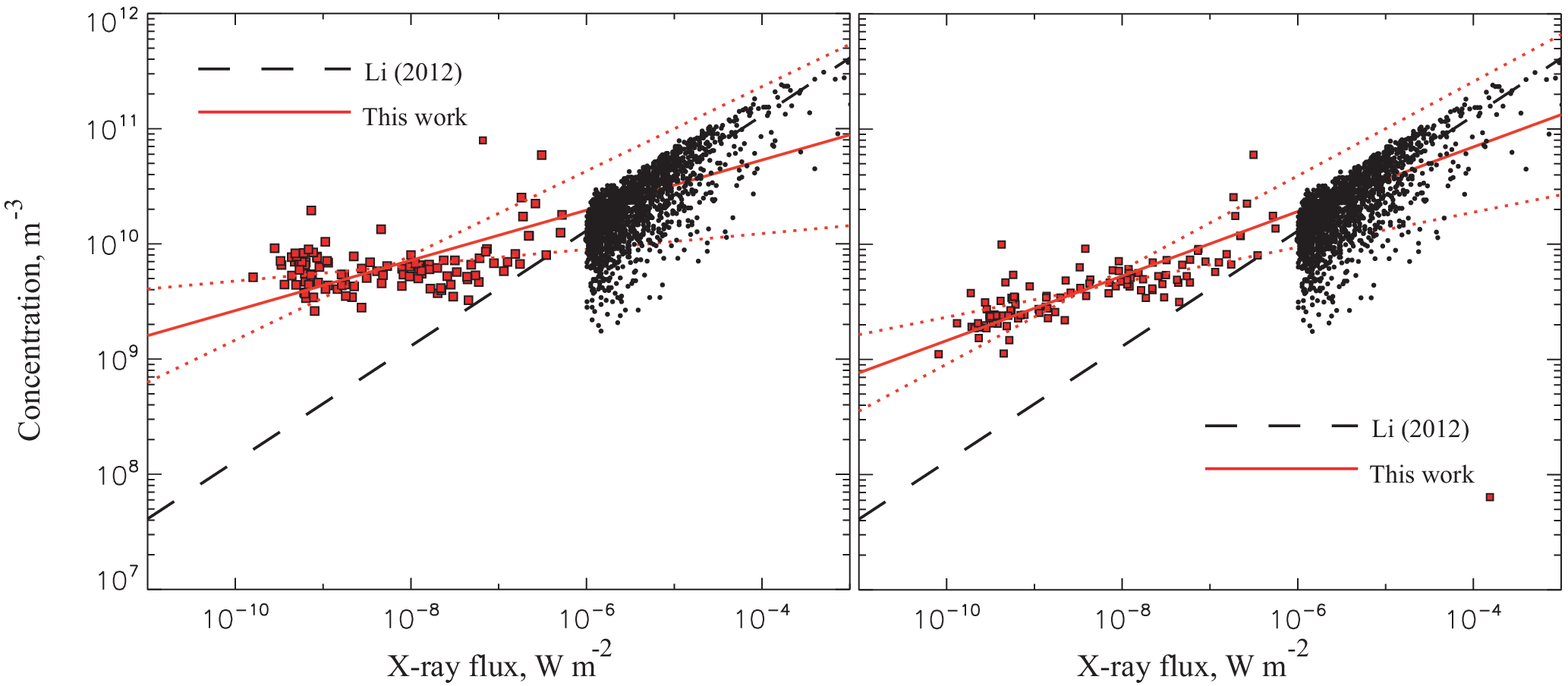}}
  \caption{Electron concentration as a function of the X-ray flux of flares within the isothermal (left) and 2T (right) models. The black circles are the data of \citet{Li2012}.}
  \label{concentration}
\end{figure*}
\subsection{The Thermal Energy and Electron Concentration}
\begin{figure*}
\centering
   \includegraphics[width=17cm]{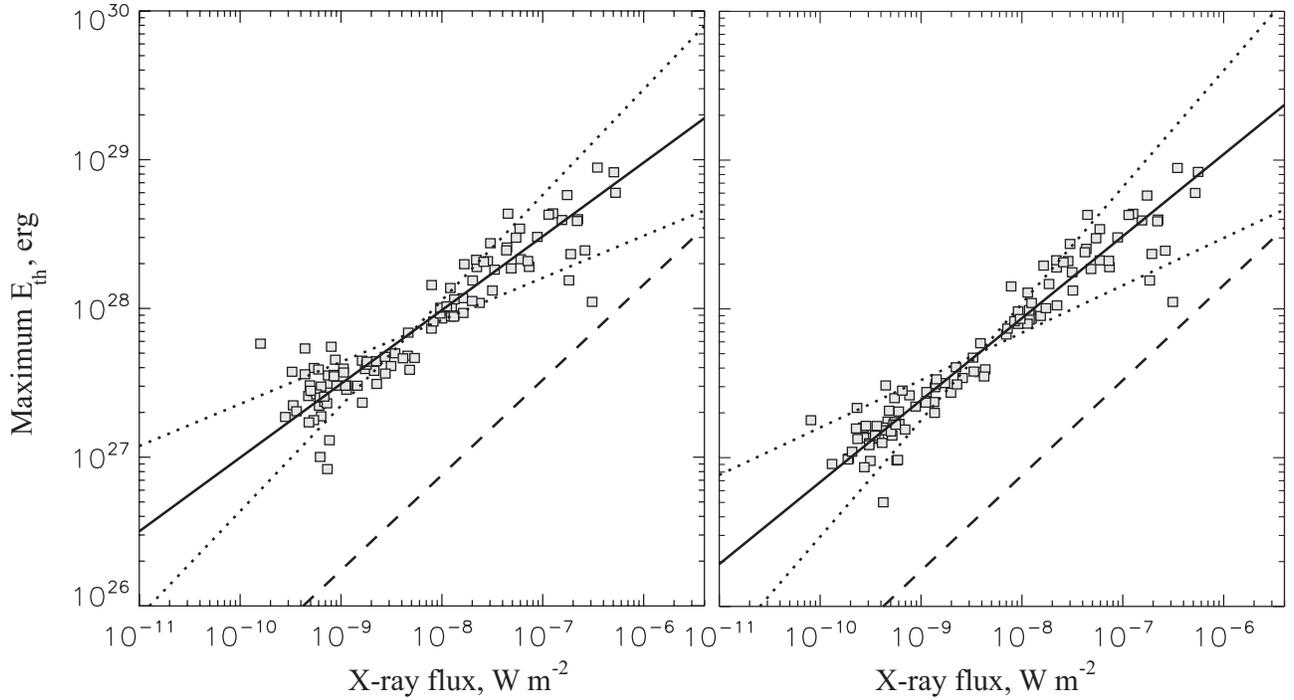}
     \caption{Thermal energy as a function of the X-ray flux of a flare, within the isothermal (left) and 2T (right) models.}
     \label{energy}
\end{figure*}

The thermal energy and electron concentration of plasma are shown in Figures~\ref{concentration} and ~\ref{energy} for the isothermal and 2T models as a function of X-ray flux. For the concentration, we compared our results with \citet{Li2012} and found that they are close for microflares of the A- and B-classes, but differ from each other for weaker events; our measurements predict a higher concentration for weak flares than could be expected from the extrapolation from the results of \citet{Li2012} in the region below A1.0. The results of fitting for this case are listed in panel V of Table~\ref{table1}. The correlation coefficients imply that the fit and experimental data are in good agreement with each other. Using these values, we obtained the thermal energy for the isothermal and 2T models (Panel VI of the Table~\ref{table1}). The slopes of the relations for the 1T and 2T models are very close to each other.\
\section{Conclusions}
We studied the efficiency of plasma heating in solar microflares as a function of their X-ray class. Earlier investigations of A-class flares and higher revealed that the plasma temperature \textit{T} and X-ray flux correlate with each other via the exponential law $log_{10}~PFF=a+bT$. However, if we extrapolate this dependence to the region of solar microflares (below A1.0 class), the predicted temperature of plasma tends to be zero (Figure~\ref{all}). Such a result seems to be in contradiction with observations of the Sun. To study this problem in detail, we performed a diagnostic of 481 solar microflares of different X-ray classes, observed with the instruments SphinX and MISH on board \textit{CORONAS-Photon}, obtained during the last solar minimum in 2009. Some of them had 100 times less X-ray emission than flares of the A1.0 class. All these events were studied in one- and two-temperature approximations. Within the 2T mode, we assumed that one of the components (the lower temperature one) corresponds to the background emission, including the emission of the quiet Sun and the active region, while another (the higher temperature one) is the emission of a flaring plasma. We did not subtract the preflare background because of some difficulties in its determination. Moreover, \citet{Ryan2012} demonstrated that subtracting the preflare background level distorts the real signal of the event. For the second component in the 2T model, we considered it as the more useful signal, corresponding to the flaring processes. The 1T model is in general less reliable than 2T because in this case we have to mix the emission from the solar flare with the background emission from the active region, which can significantly affect the results.\

The relation between $log_{10}~PFF$ and the plasma temperature that we found for microflares is different from what can be expected from observations for ordinary flares. The measured temperature of plasma for most microflares is higher than what can be predicted by simple extrapolation of the data for ordinary flares. The results obtained with the 1T and 2T models are similar but the temperature of the flaring plasma for the 2T model is higher than that for the isothermal one.\

Our measurements for flares of the A1.0 class and higher correspond well with other works, except for \citet{Battaglia2005}. However, the results obtained by \citet{Battaglia2005} strongly differ from all other works due to their specific analysis based on the \textit{RHESSI} data. The temperatures obtained from \textit{RHESSI} data are always higher than the temperatures from \textit{GOES}. Taking into account the energy range of SphinX, our results should be closer to the other papers, where \textit{GOES} soft X-ray flux was analyzed.\

For flares of the A1.0 class and higher, the relationship between plasma temperature and emission flux in the 1--8~\AA\ spectral range can be approximated by an exponential function. This is in good agreement with previous studies. However, the joint distribution, which includes normal flares ($>A1.0$) and microflares ($<A1.0$), cannot be fit by exponential. The only way to fit it this way is to use two exponential functions for flares below A1.0 and above A1.0. The best approximation for joint distribution is a power law, which is in good agreement with the data of observations in the whole flare range, from A0.01-class to X-class.\

In general, the ratio between the hot and cold components of emission increases with the X-ray flux of events. In the microflares of the B- and partially the A-classes, the X-ray flux from the hot component of plasma is 2--4 orders higher than that from the cold one. In the weakest microflares with an X-ray class 1--2 orders lower than the A1.0 level, the emission of the cold component is comparable to the hot one. This fact strongly complicates the diagnostics of hot plasma for such events. Taking this into account, we used MISH data to additionally verify the results of spectral diagnostics. As was mentioned in section~\ref{sec:data}, MISH is sensitive only to the emission of plasma with temperatures higher than 4 MK. Only 6 of 113 flares did not show any signals in the MISH data. This observational fact is in good agreement with the results of the diagnostics. The calculated temperatures for these events were in the range of 2.4--3.3 MK, which are too low to confidently register their emission with the MISH instrument.\

It should also be noted that for some of the microflares registered by MISH, the temperature of the hot component was calculated to be lower than 3.5--4 MK. To explain this we consider the fact that part of the temperature distribution in these flares was above the 4 MK level, and only the average temperature was lower than 4 MK.\

We also analyzed the correlation between the X-ray flux of the flare and the ratio \textit{T}$_{2}$/\textit{T}$_{1}$, where \textit{T}$_{2}$ and \textit{T}$_{1}$ are the temperatures of the hot and cold components of plasma within the 2T approximations (Figure~\ref{t2/t1}). With certain X-ray classes of flare, the temperatures of the heated flaring plasma \textit{T}$_{2}$ and the temperature of the surrounding plasma \textit{T}$_{1}$ become equal. This means that a flare with an X-ray class below this level cannot effectively heat the plasma. For this boundary our study gives a value of A0.0002 which corresponds to a temperature of about 1.66 MK.\

We found a correlation between the thermal energy of the heated plasma and the X-ray flux of a flare at the moment of flare maximum for the isothermal and 2T models (Figure~\ref{energy}). These results can be compared to the theoretical models, for example, Rosner-Tucker-Vaiana (RTV) scaling laws \cite{Rosner1978}. RTV uses approximations of constant pressure and uniform heating, which we believe should be well realized in the small hot loops observed in microflares. One more assumption of RTV is the stationary condition of energy balance. Due to the RTV scaling laws \cite{Aschwanden2004} $n_{e} \backsim l^{3}$, $T_{e} \backsim l^{2}$, $E_{\rm th} \backsim l^{7.5}$, where $l$ is a loop length. Taking into account these equations and assuming $V \backsim l^{3}$ one can obtain
\begin{equation}
EM=n_{e}^{2}V \backsim T_{e}^{4.5},
\label{EM_T}
\end{equation}
\begin{equation}
E_{\rm th} \backsim T_{e}^{3.75},
\label{E_T}
\end{equation}
On the other hand, from relations PFF(EM) and PFF$(T)$ (panels II, III, and VI in the Table~\ref{table1}) for the 2T model we can obtain
\begin{equation}
EM \backsim T_{e}^{4.38\pm0.88},
\label{EM_T_experiment}
\end{equation}
\begin{equation}
E_{\rm th} \backsim T_{e}^{3.95\pm0.79},
\label{E_T_experiment}
\end{equation}
Our results are close to the relationship predicted by RTV scaling laws, which supports the fact that RTV assumptions are well realized in solar microflares.\

\begin{table*}
\caption{Fitting results}
\label{table1}
\centering
\begin{tabular}{c c c c c c}
\hline\hline
\multicolumn{6}{c}{Panel I} \\
\hline
Relation & Model & a & b & LCC & $\tau$ \\
\hline
$ log_{10}~PFF=a+b~T$,~below A1.0 & 1T & -11.07$\pm$0.6 & 0.64$\pm$0.18 & 0.84 & 0.68 \\
$ log_{10}~PFF=a+b~T$,~above A1.0 & 1T & -8.93$\pm$0.66 & 0.25$\pm$0.09 & 0.92 & 0.74 \\
$ log_{10}~PFF=a+b~T$,~below A1.0 & 2T & -12.23$\pm$0.9 & 0.78$\pm$0.22 & 0.79 & 0.63 \\
$ log_{10}~PFF=a+b~T$,~above A1.0 & 2T & -8.98$\pm$0.68 & 0.25$\pm$0.09 & 0.92 & 0.74 \\
\hline\hline
\multicolumn{6}{c}{Panel II} \\
\hline
$ log_{10}~PFF=a+b~log_{10}EM $ & 1T & -122.19$\pm$25.07 & 2.46$\pm$0.54 & 0.79 & 0.62 \\
$ log_{10}~PFF=a+b~log_{10}EM $ & 2T & -83.75$\pm$14.9 & 1.64$\pm$0.32 & 0.96 &  0.82 \\
\hline\hline
\multicolumn{6}{c}{Panel III} \\
\hline
$ log_{10}~PFF=a+b~log_{10}T $ & 1T & -11.54$\pm$0.67 & 5.16$\pm$1.02 & 0.96 & 0.84 \\
$ log_{10}~PFF=a+b~log_{10}T $ & 2T & -13.29$\pm$1 & 7.18$\pm$1.45 & 0.9 & 0.8 \\
\hline\hline
\multicolumn{6}{c}{Panel IV} \\
\hline
$ log_{10}~PFF=a+b~log_{10}({T_{2}}/{T_{1}})  $ & 2T & -11.71$\pm$0.71 & 6.84$\pm$1.9 & 0.9 & 0.74 \\
\hline\hline
\multicolumn{6}{c}{Panel V} \\
\hline
$ log_{10}N_{e}=a+b~log_{10}~PFF $ & 1T & 11.6$\pm$1.23 & 0.22$\pm$0.15 & 0.36 & 0.12 \\
$ log_{10}N_{e}=a+b~log_{10}~PFF $ & 2T & 11.96$\pm$1.08 & 0.28$\pm$0.13 & 0.81 & 0.65\\
\hline\hline
\multicolumn{6}{c}{Panel VI} \\
\hline
$ log_{10}E_{th}=a+b~log_{10}~PFF $ & 1T & 31.96$\pm$1.77 & 0.5$\pm$0.21 & 0.92 & 0.74 \\
$ log_{10}E_{th}=a+b~log_{10}~PFF $ & 2T & 32.35$\pm$1.96 & 0.55$\pm$0.23 & 0.96 & 0.82 \\
\hline
\end{tabular}
\end{table*}

\begin{acknowledgements}
This work was supported by the Russian Science Foundation (RSF) grant No. 17-12-01567. We also thank the anonymous reviewer for revision of the work and for useful comments that improved the manuscript.
\end{acknowledgements}
\newpage
\bibliography{lit}

\begin{thebibliography}{}
\expandafter\ifx\csname natexlab\endcsname\relax\def\natexlab#1{#1}\fi

\bibitem[{{Aschwanden}(2004)}]{Aschwanden2004}
{Aschwanden}, M.~J. 2004, {Physics of the Solar Corona. An Introduction}
  (Praxis Publishing Ltd)

\bibitem[{{Battaglia} {et~al.}(2005){Battaglia}, {Grigis}, \&
  {Benz}}]{Battaglia2005}
{Battaglia}, M., {Grigis}, P.~C., \& {Benz}, A.~O. 2005, \aap, 439, 737

\bibitem[{{Caspi} {et~al.}(2014){Caspi}, {Krucker}, \& {Lin}}]{Caspi2014}
{Caspi}, A., {Krucker}, S., \& {Lin}, R.~P. 2014, \apj, 781, 43

\bibitem[{{Feldman} {et~al.}(1996){Feldman}, {Doschek}, {Behring}, \&
  {Phillips}}]{Feldman1996}
{Feldman}, U., {Doschek}, G.~A., {Behring}, W.~E., \& {Phillips}, K.~J.~H.
  1996, \apj, 460, 1034

\bibitem[{{Garcia} \& {McIntosh}(1992)}]{Garcia1992}
{Garcia}, H.~A., \& {McIntosh}, P.~S. 1992, \solphys, 141, 109

\bibitem[{{Hannah} {et~al.}(2008){Hannah}, {Christe}, {Krucker}, {Hurford},
  {Hudson}, \& {Lin}}]{Hannah2008}
{Hannah}, I.~G., {Christe}, S., {Krucker}, S., {et~al.} 2008, \apj, 677, 704

\bibitem[{{Kirichenko} \& {Bogachev}(2013)}]{Kirichenko2013}
{Kirichenko}, A.~S., \& {Bogachev}, S.~A. 2013, Astronomy Letters, 39, 797

\bibitem[{{Kuzin} {et~al.}(2009){Kuzin}, {Bogachev}, {Zhitnik}, {Pertsov},
  {Ignatiev}, {Mitrofanov}, {Slemzin}, {Shestov}, {Sukhodrev}, \&
  {Bugaenko}}]{Kuzin2009}
{Kuzin}, S.~V., {Bogachev}, S.~A., {Zhitnik}, I.~A., {et~al.} 2009, Advances in
  Space Research, 43, 1001

\bibitem[{{Li} {et~al.}(2012){Li}, {Gan}, \& {Feng}}]{Li2012}
{Li}, Y.~P., {Gan}, W.~Q., \& {Feng}, L. 2012, \apj, 747, 133

\bibitem[{{Mrozek} {et~al.}(2012){Mrozek}, {Gburek}, {Siarkowski}, {Sylwester},
  {Sylwester}, \& {Gryciuk}}]{Mrozek2012}
{Mrozek}, T., {Gburek}, S., {Siarkowski}, M., {et~al.} 2012, Central European
  Astrophysical Bulletin, 36, 71

\bibitem[{{Rosner} {et~al.}(1978){Rosner}, {Tucker}, \& {Vaiana}}]{Rosner1978}
{Rosner}, R., {Tucker}, W.~H., \& {Vaiana}, G.~S. 1978, \apj, 220, 643

\bibitem[{{Ryan} {et~al.}(2012){Ryan}, {Milligan}, {Gallagher}, {Dennis},
  {Tolbert}, {Schwartz}, \& {Young}}]{Ryan2012}
{Ryan}, D.~F., {Milligan}, R.~O., {Gallagher}, P.~T., {et~al.} 2012, \apjs,
  202, 11

\bibitem[{{Sylwester} {et~al.}(2008){Sylwester}, {Kuzin}, {Kotov}, {Farnik}, \&
  {Reale}}]{Sylwester2008}
{Sylwester}, J., {Kuzin}, S., {Kotov}, Y.~D., {Farnik}, F., \& {Reale}, F.
  2008, Journal of Astrophysics and Astronomy, 29, 339

\end{thebibliography}
\end{document}